\documentclass[fleqn,twoside]{article}
\usepackage{espcrc2}

\usepackage{graphicx}

\newcommand{\AmS}{{\protect\the\textfont2
  A\kern-.1667em\lower.5ex\hbox{M}\kern-.125emS}}

\title{Analytic structure of the gluon and
       quark propagators in Landau gauge QCD \thanks{Summary of a 
        talk given at several occasions; 
	to be published in the proceedings of the international 
	conference QCD DOWN UNDER, March 10 - 19, Adelaide, Australia}
	}

\author{R.~Alkofer, {Institute for Theoretical Physics, University of
       T\"ubingen, D-72076 T\"ubingen, Germany,}\\
        W.~Detmold, {Department of Physics, University of Washington,
             Box  351560, Seattle WA 98195, USA},\\
	C.~S.~Fischer, {IPPP, University of Durham, Durham DH1 3LE, U.K.},\\
	P.~Maris, {U.\ of Pittsburgh, Dept.\ of Physics and Astronomy,
	100 Allen Hall, 
	Pittsburgh PA 15260, USA}
	}
       
\begin{document}

\begin{abstract}
  In Landau gauge QCD the infrared behavior of the propagator of
  transverse gluons can be analytically determined to be a power law
  from Dyson--Schwinger equations. This propagator clearly shows
  positivity violation, indicating the absence of the transverse
  gluons from the physical spectrum, {\it i.e.\/} gluon confinement.
  A simple analytic structure for the gluon propagator is proposed
  capturing all important features. We provide arguments that the
  Landau gauge quark propagator possesses a singularity on the real
  timelike axis.  For this propagator we find a positive definite
  Schwinger function.
\vspace{1pc}
\end{abstract}

\maketitle


The standard model of particle physics consists of gauge field
theories. These have been postulated on the basis of symmetries and
their elementary excitations do not reflect the observed particle
spectrum. In the quantum formulation of these theories, especially in
Poincar{\'e}-covariant gauges, an intricate problem is posed by the
separation of physical and unphysical degrees of freedom.

In Quantum Electrodynamics (QED) in linear covariant gauges, the
electromagnetic field can be decomposed into transverse, longitudinal
and timelike photons, however, only transverse polarizations are
observed. From a purely mathematical point of view, this can be
understood from the representations of the Poincar{\'e} group for
massless states: massless particles have only two possible
polarizations \cite{Wigner56}. The apparent contradiction is resolved
by the fact that timelike and longitudinal photons cancel exactly in
the $S$-matrix \cite{Bleuler50}. In this context we emphasize that the
states of quantum gauge field theories in covariant gauges necessarily
constitute an indefinite metric space. In covariant gauge QED one thus
has to sacrifice the principle of positivity of the representation
space.

In Quantum Chromodynamics (QCD) in linear covariant gauges, the
cancellation of unphysical degrees of freedom in the $S$-matrix is
substantially complicated by the self-interaction of the gauge fields
and by the ghost fields that are necessarily present in the quantum
formulation of these theories \cite{Faddeev67}. To order $\alpha_S^2$
in perturbation theory, one obtains amplitudes for the scattering of
two transverse gluons into one transverse and one longitudinal gluon.
However, at the same order, a ghost loop appears and cancels the
various gluon loops, and scattering of transverse to longitudinal
gluons does not occur. It is possible to prove this cancellation to
all orders in perturbation theory on the basis of the BRS
\cite{Becchi:1976nq} symmetry of the covariantly gauge fixed theory.
This symmetry can be represented by gauge transformations with the
ghost field as a parameter.  The ghost fields, being scalar,
anti-commuting fields, are necessarily in the unphysical part of the
representation space. In covariant gauge QCD, the physical (and thus
positive definite) part of the state space is conjectured to be the set of
BRS singlets \cite{Nakanishi:qm}.

Gauge fixed QCD is invariant under transformations related to the
ghost number, and employing such transformations, one can show that
BRS non-singlets occur in quartets \cite{Nakanishi:qm}. Such a
BRS-quartet consists of two parent and two daughter states of
respectively opposite ghost numbers. The latter states are BRS-exact
and thus BRS-closed because the BRS transformation is nilpotent. The
BRS daughters are orthogonal to all other states in the positive
definite subspace and therefore do not contribute to physical
$S$-matrix elements.  The parent states belong to the indefinite
metric part of the representation space and are expected to violate
positivity.

The Kugo--Ojima confinement scenario \cite{Kugo:1979gm,Nakanishi:qm}
describes a mechanism by which the positive semi-definite, physical
state space contains only colorless states. Colored states are not
BRS singlets and therefore do not appear in $S$-matrix elements: they
are confined.  Thus an investigation of (non-)positivity of transverse
gluons and quarks allows us to understand confinement
via the BRS quartet mechanism in more detail.

Here we present analytic properties of the gluon and
quark propagators in Landau gauge QCD as they result from
non-perturbative calculations (more details can be found 
in Ref.\ \cite{Paper}).
We confirm previous results
\cite{vonSmekal:1997is,Mandula:nj} on positivity violation for the
gluon propagator.  We also provide a parameterization of the gluon
propagator that is analytic throughout the complex $p^2$ plane except
on the real timelike axis and which decreases to zero in every
direction of the complex $p^2$ plane. Such behavior satisfies the
usual axioms of local quantum field theory \cite{Oehme:1994pv} (except
positivity).
For the quark propagator, we analyze several general constraints,
lattice data \cite{LATTICE_QUARK}, and solutions of the coupled
quark-gluon-ghost Dyson--Schwinger equations (DSEs)
\cite{Fischer:2003rp}.  None of these contradict positivity of the
quark propagator.

Within the framework of a Euclidean quantum field theory (used
in the following), positivity is formulated in terms of the
Osterwalder--Schrader axiom of reflection positivity
\cite{Osterwalder:1973dx}.  In the special case of a two point
correlation function, $\Delta(x-y)$, the condition of reflection
positivity can be written as
\begin{equation}
\int d^4x \; d^4y \;
\bar{f}(\vec{x},-x_0) \; \Delta(x-y) \; {f}(\vec{y},y_0) \; \ge 0 \;,
\end{equation}
where $f(\vec{x},x_0)$ is a complex valued test function with support in
$\{(\vec{x},x_0) \: : \: x_0 > 0 \}$.
After a three-dimensional Fourier transformation,
this condition can be given in terms of the {Schwinger function} 
\begin{eqnarray}
  \Delta(t)  &=& \frac{1}{\pi}\int
  dp \cos(t p) \sigma(p^2) \;\ge 0 \,,
\label{schwinger}
\end{eqnarray}
where $\sigma(p^2)$ is a scalar function extracted from the
corresponding two-point function.

The elementary two-point functions of QCD are the ghost, gluon, and
quark propagators.  In Landau gauge these renormalized momentum-space
propagators $D_G(p)$, $D_{\mu \nu}(p)$, and $S(p)$ can be generically
written as
\begin{eqnarray}
  D_G(p) &=& - \frac{G(p^2)}{p^2} \,,
  \label{ghost_prop}\\
  D_{\mu \nu}(p) &=& \left(\delta_{\mu \nu} - \frac{p_\mu
      p_\nu}{p^2} \right) \frac{Z(p^2)}{p^2} \, ,
  \label{gluon_prop} \\  
  S(p) &=& \frac{1}{-i  p\!\!\!/\, A(p^2) + B(p^2)} \nonumber \\
 &=&  i p\!\!\!/\, \sigma_v(p^2) + \sigma_s(p^2)\,.\hspace*{5mm}
  \label{quark_prop}
\end{eqnarray}
Note that the quark propagator $S(p)$ is decomposed into scalar and
vector parts, $\sigma_{s}(p^2)$ and $\sigma_{v}(p^2)$.  Violation of
reflection positivity can be studied by calculating
Eq.~(\ref{schwinger}) with $\sigma_{g}(p^2)= Z(p^2)/p^2$ for the
transverse gluons and $\sigma_{s}(p^2)$ and $\sigma_{v}(p^2)$ for the
quarks.  (The ghost propagator violates reflection positivity by the
way ghosts are introduced in Faddeev--Popov quantization.)

The coupled set of DSEs (for recent reviews see {\it e.g.\/}
\cite{Alkofer:2000wg}) for the ghost and gluon propagators 
can be solved analytically for $p^2 \to 0^+$ \cite{Watson:2001yv}.
One finds simple power laws for the gluon and ghost dressing functions
\begin{eqnarray}
  Z(p^2) \sim (p^2)^{2\kappa}, \qquad
  G(p^2) \sim (p^2)^{-\kappa}.
  \label{g-power}
\end{eqnarray}
such that the product $Z(k^2)G^2(k^2)\sim \alpha_S(k^2)$ goes to a
constant in the infrared: there is an infrared fixed point for the
running coupling.  The value of the exponent $\kappa$ is in the range
$0.5 < \kappa < 0.7$, depending on details of the truncation of the
set of DSEs
\cite{Fischer:2003rp,Fischer:2002hn,Lerche:2002ep,Zwanziger:2001kw}.
Here we use a self-consistent truncation scheme
\cite{Fischer:2003rp,Fischer:2002hn} which neglects the effects of the
four-gluon interaction and employs {\it ans\"atze} for the
ghost-gluon, quark-gluon, and three-gluon vertices such that two
important constraints are fulfilled: the running coupling,
$\alpha_S(p^2)$, is independent of the renormalization point, and the
anomalous dimensions of the propagators are reproduced at one-loop
level for large momenta.  In this particular truncation $\kappa = (93
- \sqrt{1201})/98 \approx 0.595$ \cite{Lerche:2002ep,Zwanziger:2001kw}
is an irrational number, and $\alpha_S(0) \approx 2.972$.  Infrared
dominance of the gauge fixing part of the QCD action
\cite{Zwanziger:2003cf} implies infrared dominance of ghosts in the
DSEs which, in turn, can be used to show \cite{Lerche:2002ep} that
$\alpha_S(0)$ depends only weakly on the dressing of the ghost-gluon
vertex and not at all on other vertex functions.  Recently, the same
infrared behavior of the propagators has been found using the method
of Exact Renormalization Group Equations \cite{Pawlowski:2003hq}.

\begin{figure}
\includegraphics[width=\columnwidth]{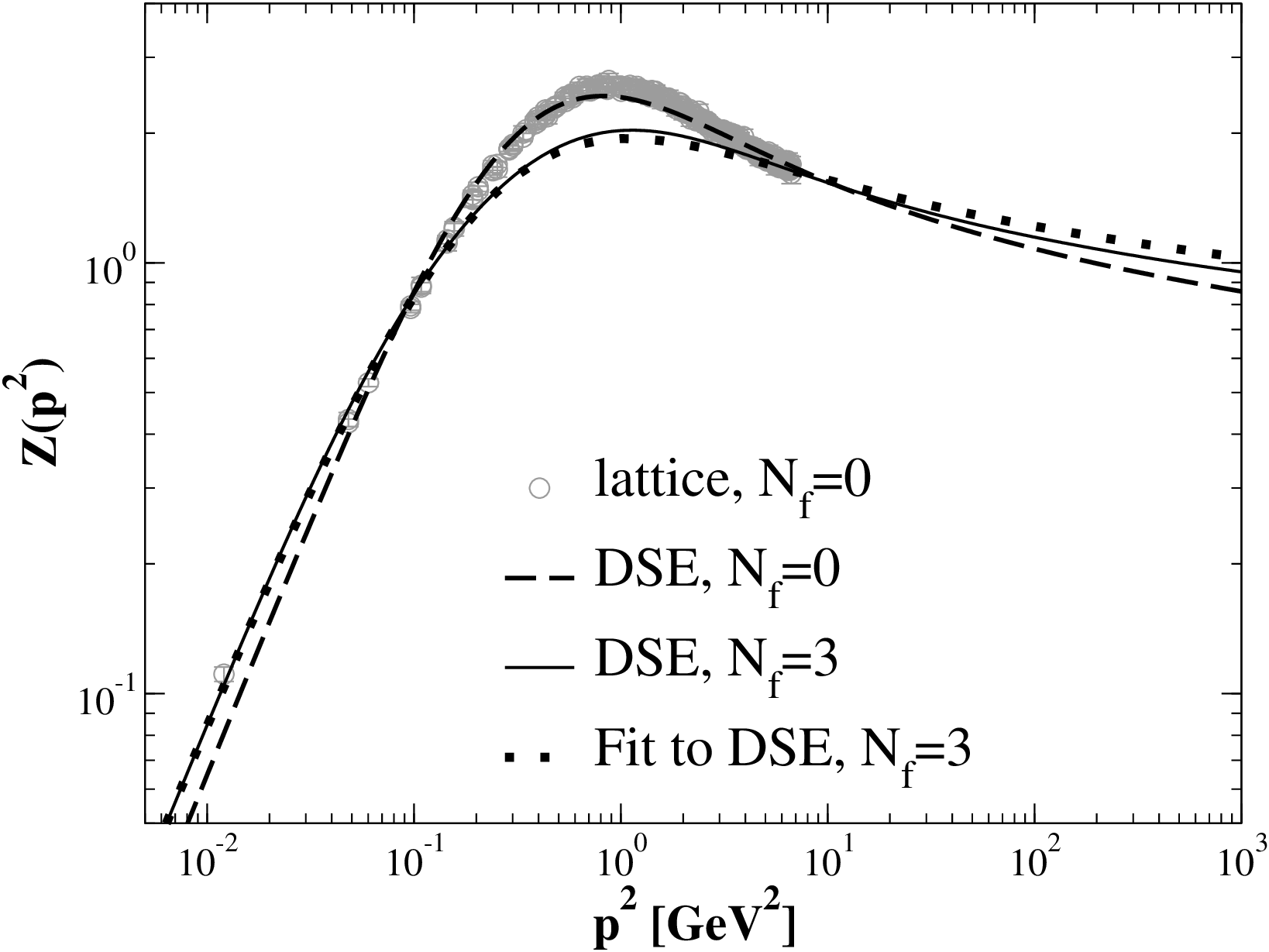}
\caption{DSE \protect\cite{Fischer:2003rp}
and lattice \protect\cite{Bonnet:2001uh} results for the gluon renormalization 
  function $Z(p^2)$ and the fit (\protect{\ref{fitII}}) are shown.}
\label{XX1}
\end{figure}

For quenched QCD, the gluon propagator, as it results from numerical
solution of the coupled DSEs for the gluon and ghost propagators,
agrees very well with recent lattice data \cite{Bonnet:2001uh}, see
Fig.~\ref{XX1}.  The unquenched DSE gluon propagator is significantly
suppressed in the intermediate momentum region, where screening effects
of $q\bar{q}$ pairs becomes important.  For both $N_f=0$ and $N_f=3$
the infrared behavior is given by Eq.~(\ref{g-power}).

The corresponding running coupling can be accurately represented by
\cite{Fischer:2003rp}
\begin{eqnarray}
&&\alpha_{\rm fit}(p^2) = 
\nonumber\\
&&\frac{1}{1+(p^2/\Lambda^2_{\tt QCD})}
\Bigg[\alpha_S(0) + (p^2/\Lambda^2_{\tt QCD}) \times \nonumber\\
&& \hspace*{3mm}  
\frac{4 \pi}{\beta_0}
\left(\frac{1}{\ln(p^2/\Lambda^2_{\tt QCD})} 
- \frac{1}{p^2/\Lambda_{\tt QCD}^2 -1}\right)\Bigg] \, , \hspace*{3mm} 
\label{fitB}
\end{eqnarray}
with $\beta_0=(11N_c-2N_f)/3$.  Note that the Landau pole at spacelike
$p^2=\Lambda^2_{\tt QCD}$ is canceled, {\it c.f.\/} Ref.\ 
\cite{Shirkov:1997wi}.  The expression (\ref{fitB}) is analytic in the
complex $p^2$ plane except on the real timelike axis ($p^2<0$) where
the logarithm produces a cut.

Since the infrared exponent, $\kappa$, is an irrational number, one
knows that the corresponding gluon propagator possesses a cut on the
negative real axis as well. It is possible to fit the DSE solution for
the gluon propagator without introducing further singularities.  The
fit to the gluon renormalization function with the form
\begin{equation}
Z_{\rm fit}(p^2) = w \left(\frac{p^2}{\Lambda^2_{\tt QCD}+p^2}\right)^{2 \kappa}
 \left( \alpha_{\rm fit}(p^2) \right)^{-\gamma}\,,
 \label{fitII}
\end{equation}
with $w= 2.5$ and $\Lambda_{\tt QCD}=510$ MeV, is in good
agreement with the DSE solution.  Here $w$ is a normalization
parameter, and $\gamma = (-13 N_c + 4 N_f)/(22 N_c - 4 N_f)$ is the
one-loop value for the anomalous dimension of the gluon propagator.
The discontinuity across the cut in $Z_{\rm fit}(p^2)$ vanishes for
$p^2\to 0^-$, diverges to $+\infty$ at $p^2=-\Lambda_{\tt QCD}^2$ on
both sides and drops to zero for $p^2\to - \infty$.  Additional
parameterizations have been explored in Ref.\ \cite{Paper}.

\begin{figure}
\includegraphics[width=\columnwidth]{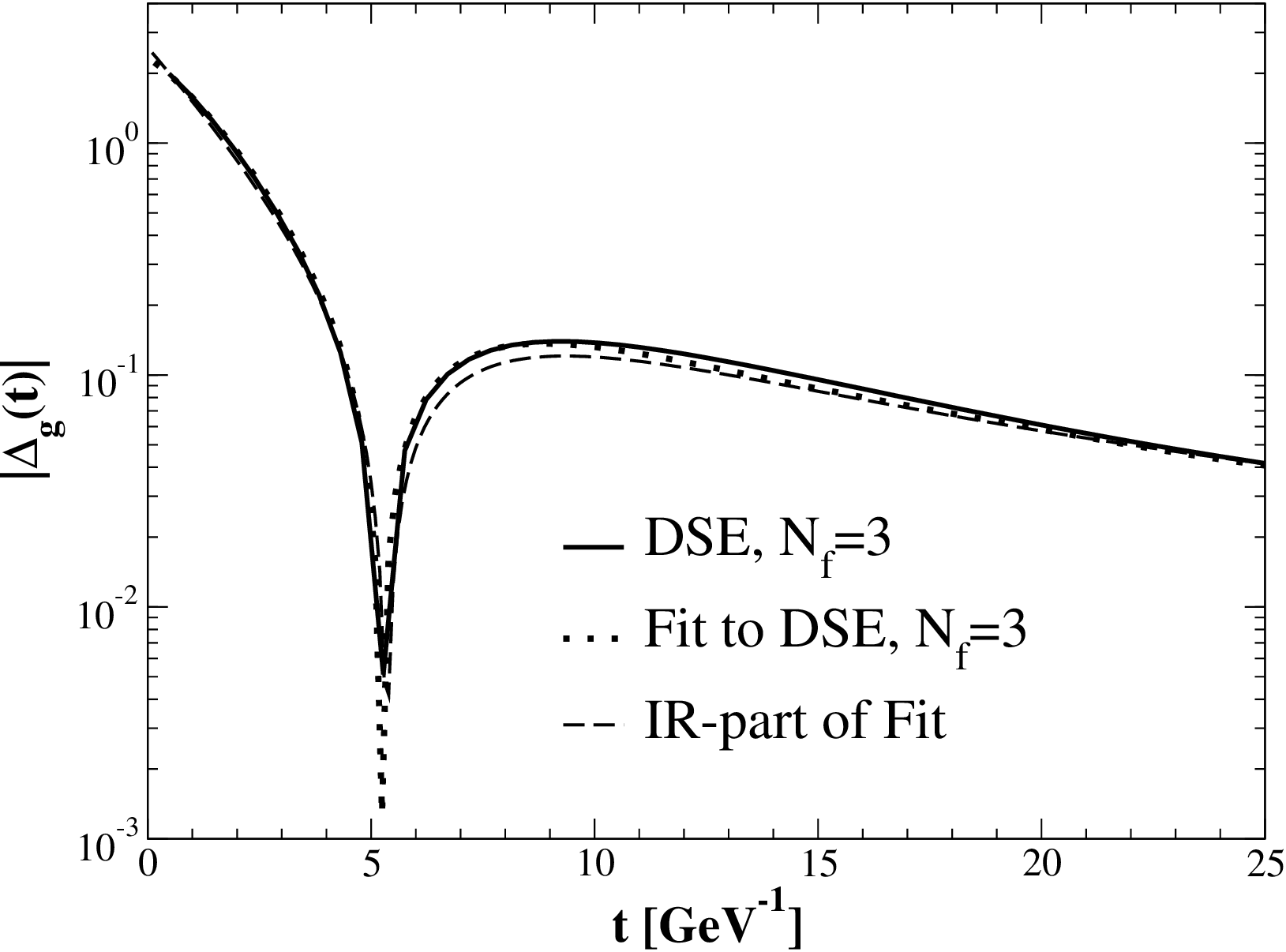}
\caption{Shown are the results for the absolute value of the gluon
Schwinger function from the DSE solution, the fit (\protect{\ref{fitII}}),
and the infrared part of this fit.}
\label{XX2}
\end{figure}

The corresponding Schwinger function, $\Delta_g (t)$, based on the fit,
Eq.~(\ref{fitII}), is compared to the DSE solution in Fig.~\ref{XX2}. 
To enable a logarithmic scale, the absolute value is displayed.
$\Delta_g (t)$ has a zero for 
$t\approx 5 \, {\rm GeV}^{-1} \approx  1 \, {\rm fm}$ 
and is negative for larger Euclidean times. {\it I.e.\/} we
clearly observe positivity violations in the DSE gluon propagator and
the agreement of the numerical Schwinger function with the Fourier
transformed fit is also excellent. The crucial property for positivity
violation in the gluon propagator is that it vanishes for $p^2\to 0^+$.
This can be seen from the relation $ 0 =
\sigma_g(p^2=0) = \int {d^4x} \; D(x)$ which implies that the gluon
propagator in coordinate space, $D(x)$, is trivially zero or
necessarily contains positive as well as negative contributions.

\goodbreak

In expression (\ref{fitII}), the overall magnitude, $w$, is arbitrary
because of renormalization properties (it is determined via the choice
of the renormalization scale).  The infrared exponent, $\kappa$, is
determined from the infrared analysis of the DSEs, and the one-loop
value is used for the gluon anomalous dimension, $\gamma$. Thus,  
we have found a parameterization of the gluon propagator
which has effectively only one parameter, the scale $\Lambda_{\tt QCD}$.
This and the relatively simple analytic structure gives us confidence
that we have succeeded in uncovering the important features of the
Landau gauge gluon propagator.

\begin{figure}
\includegraphics[width=\columnwidth]{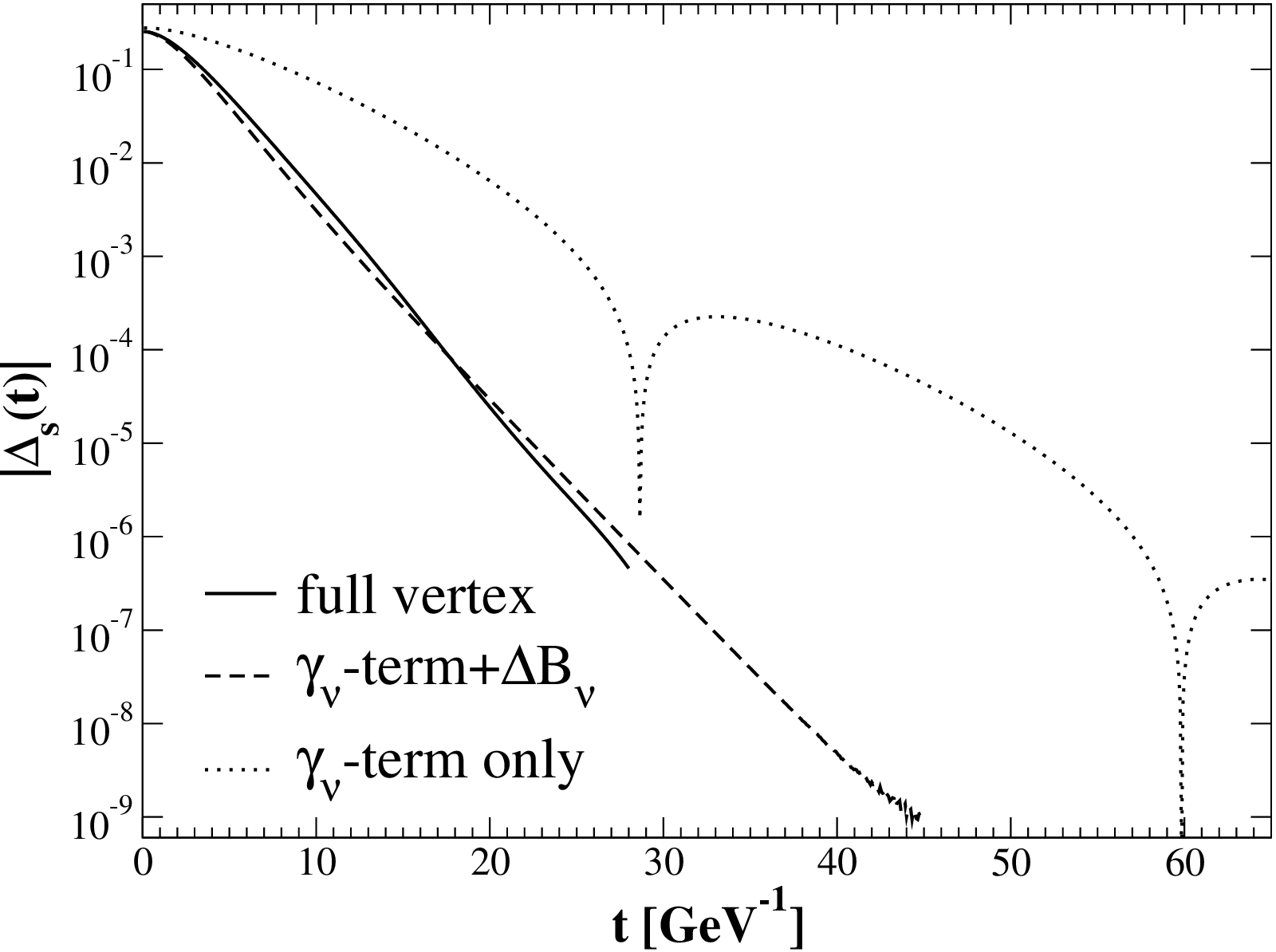}
\caption{Results for the quark Schwinger functions using different
  approximations to the quark-gluon vertex.}
\label{XX3}
\end{figure}

We now turn to the analytic structure of the quark propagator.  In the
DSE studies of Ref.\ \cite{Fischer:2003rp} it has been assumed that
the non-Abelian part of the quark-gluon vertex can be factored out
from the tensor structure. This structure is then given by the
Curtis--Pennington vertex \cite{Curtis:1990zs} which, by construction,
is multiplicatively renormalizable and satisfies the Abelian
Ward--Takahashi identity of QED. In this particular case one has a term
\begin{equation}
  \Delta B_\nu := i\frac{B(p^2)-B(q^2)}{p^2-q^2} (p+q)_\nu\, ,
  \label{DeltaB}
\end{equation}
in the vertex.  For the present discussion it is important to note
that the exact quark-gluon vertex (which satisfies the more
complicated Slavnov--Taylor identity of QCD), and future improved
approximations to it, will almost certainly also exhibit such a
quark-gluon coupling proportional to the sum of quark momenta. This
coupling, being effectively scalar, is not invariant under chiral
transformations in contrast to the leading $\gamma_\nu$ part of the
vertex.  However, the term in Eq.~(\ref{DeltaB}) only appears once
chiral symmetry is broken dynamically and is thus consistent with the
chiral Ward identities.  Its existence provides a significant
self-consistent enhancement of dynamical chiral symmetry breaking in
the quark DSE, which is necessary to produce an acceptable value of
the chiral condensate \cite{Fischer:2003rp}.  Quite independently of
the form of the gluon propagator, the resulting quark propagator
respects positivity if the term in Eq.~(\ref{DeltaB}) is included in
the quark-gluon vertex \cite{Paper}, as evidenced by the Schwinger
functions in Fig.~\ref{XX3} (solid and dashed curves). From this
figure we also see that positivity is violated if only the bare
quark-gluon vertex is used (dotted curve), in agreement with previous
work employing this approximation~\cite{ccsings}.

A single real pole on the negative momentum axis results in a pure
exponential decay of the corresponding Schwinger function.  This is
not what is observed in $\Delta_s(t)$ for small times where there is
some curvature (see Fig.~\ref{XX3}). To give an acceptable
reproduction of $\Delta_s(t)$, analytic parameterizations of the quark
propagator must contain additional sub-dominant structure.  By Fourier
transforming combinations of functions with cuts and poles at timelike
momenta we have determined three simple sources for such curvature.
The dominant singularity may be accompanied by additional real
singularities at larger mass scales, or by complex conjugate
singularities with a larger real part of the mass, or it may be the
starting point of a branch cut on the negative real momentum axis
\cite{Paper}. Of course such a list is by no means exhaustive.

In Ref.~\cite{Paper}, we have used various general requirements,
lattice data \cite{LATTICE_QUARK}, and the DSE solutions
\cite{Fischer:2003rp} to constrain these various parametric
forms. Such analysis leads to an important conclusion: in all
parameterizations, the dominant singularity must occur on (or {\it
very} close to) the real, timelike half-axis at a scale $m\sim 350$ to
$500$~MeV, the lower estimate coming when lattice data are used for
the fit, the upper estimate for fits to the DSE solutions.  This scale
may relate to a constituent quark mass.

Whilst this result is robust, our current methods are not able to
accurately determine the nature of the dominant singularity or
reliably constrain the additional sub-dominant structures.  In
particular, the regular infrared behavior of quark propagator found in
the DSE and lattice calculations makes determining its detailed
analytic properties very difficult without explicitly probing the
timelike momentum half-plane.  A more conclusive analysis of the
structure of the quark propagator would result from direct solution of
the DSEs over an appropriate region of the complex momentum plane.

To summarize: We have proposed the relatively simple function,
Eq.~(\ref{fitII}), to describe the full (non-perturbative) Landau
gauge gluon renormalization function for all complex values of
momentum. The corresponding gluon propagator agrees well with DSE
solutions and lattice data for spacelike momenta, it is positivity
violating, and it is analytic everywhere except for a cut on the
negative real $p^2$ half-axis. Thus it implies that in Landau gauge
QCD, the confinement of transverse gluons is related to the violation
of Osterwalder--Schrader reflection positivity.  We have also provided
evidence that the Schwinger functions related to the quark propagator
are positive definite, and consequently quark confinement is not
manifest at the level of the propagator.  A number of relatively
simple parameterizations have been suggested for this propagator in
terms of real and complex conjugate poles and branch cuts. In all
cases, the dominant singularity is real and occurs at a scale $350$ to
$500$~MeV.

\medskip

\section*{Acknowledgements}

We are grateful to P.~van Baal, P.~Bowman, H.~Gies, C.~Roberts,
D.~Shirkov, I.~Solovtsov, O.~Solovtsova, P.~Tandy, A.~Williams,
J.~Zhang, and D.~Zwanziger for helpful discussions.  
W.D.\ thanks the members of the ITP of the University of T\"ubingen
for their hospitality during his visits.  We thank the ECT* for the
support of the workshop ``Aspects of Confinement and Non-perturbative
QCD'' which has been important for the success of this investigation.
This work has been supported by the Deutsche For\-schungsgemeinschaft
(DFG) under contracts Al 279/3-3, Al 279/3-4, Gi 328/1-2 and GRK683
and by the US Department of Energy contract DE-FG03-97ER41014.

Last, but not least, we thank Ayse Kizilers{\"u}, Tony Thomas and 
Tony Williams
for the organisation of this exceptional conference QCD DOWN UNDER.

\end{document}